\documentclass[article]{aastex631}

\usepackage{graphicx} % Required for inserting images
\usepackage{amsfonts} 
\usepackage{graphicx}
\usepackage[version=4]{mhchem}
\usepackage{amssymb}
\usepackage{lipsum}
\usepackage{wrapfig}
\usepackage{amsmath}
\usepackage[T1]{fontenc}
\usepackage{subfigure}

\usepackage{hhline}
\usepackage{multirow}

\begin{document}

% \title{Template \aastex Article with Examples: 
% v6.3.1\footnote{Released on March, 1st, 2021}}

\title{Characterizing the Scale Height and Filamentary Structure of Radiatively Cooled MADs}

\correspondingauthor{Akshay Singh}
\email{akshay.singh@biu.ac.il}

\author[0009-0006-7515-5164]{Akshay Singh}
\affiliation{Bar-Ilan University, Ramat-Gan 5290002, Israel. \\}

\author[0000-0003-4477-1846]{Damien B\'egu\'e }
\affiliation{Bar-Ilan University, Ramat-Gan 5290002, Israel. \\}

% \collaboration{20}{(AAS Journals Data Editors)}

\author[0000-0001-8667-0889]{Asaf Pe'er}
\affiliation{Bar-Ilan University, Ramat-Gan 5290002, Israel. \\}

\begin{abstract}
% 250 words limit
Radiative cooling can strongly influence the structure and dynamics of black hole accretion disks.
Here, we perform general relativistic magnetohydrodynamic (GR-MHD) simulations of magnetically arrested disks (MADs) around a non-spinning black hole. Radiative
cooling is consistently included in the simulations and its intensity is scaled by the mass accretion rate ranging from $10^{-7}$ to $10^{-4} \dot{M}_{\mathrm{Edd}}$. Considering synchrotron and bremsstrahlung emission, we quantify
how radiative losses modify the disk structure and the accretion dynamics.
In the inner MAD disk regions, accumulation of magnetic field regulates gas accretion, enforcing
the gas into a discrete interchange-driven filamentary structure. We identify, both analytically and numerically, a transition
mass accretion rate above which radiative cooling becomes faster than the heating, which is assumed to occur via local coupling to the magnetic field.
Above this mass accretion rate, cooling substantially reduces the gas thermal pressure, leading to considerably thinner and 
denser accretion filaments, and a substantial increase in radiative efficiency, relative to lower accretion rates. We show that under these conditions,
conventional measures of the disk scale height become misleading in MAD flows. We therefore introduce an alternative
definition based on the polar position of the density maximum, which more
robustly characterizes the filamentary structure of the disks in the presence of strong magnetic
fields and cooling.
\end{abstract}


\section{Introduction} \label{sec:intro}

The study of accretion onto black holes (BHs) is fundamental to our understanding of various objects such as X-ray emitting binaries (XRBs), active galactic
nuclei (AGN) and the evolution of galaxies. These disks are dynamic, rotating structures
made up of gas and dust, frequently in a plasma state due to ionization. In these systems, matter gradually
gather around a central object under its gravitational force. The characteristics and configuration of these
disks are determined by factors such as the rate at which mass is accreted onto the BH and the nature of the
magnetic fields, either generated within the disk itself or carried along with the accreting material from
far distances \citep[for reviews see, e.g.,][]{ Abramowicz+13, Yuan+14}. 

Within the framework of magnetohydrodynamics (MHD), two primary regimes of accretion have emerged based on the configuration and strength of the magnetic field: the Standard and Normal Evolution (SANE) state \citep{Narayan+12} and the Magnetically Arrested Disk (MAD) state \citep{Kogan+76,Narayan+003, Igu+08}. While SANE disks are characterized by weak, turbulent magnetic fields that do not significantly alter the gas dynamics \citep[albeit promoting accretion via magneto-rotational-instability, see e.g.][]{Narayan+12}, MADs accumulate enough magnetic flux in the vicinity of the BH to dynamically influence the inward flow. In this state, the magnetic pressure 
becomes comparable to the ram pressure of the infalling gas \citep{Tchekho+12,Narayan+22}. This leads to a regulated accretion process in which the gas no longer flows continuously, but instead proceeds through discrete blobs or streams that penetrate the magnetosphere via the magnetic interchange instability \citep{Lubow+94, Ripperda+22}.

Both SANE and MAD regimes were extensively studied numerically in the past two decades, using general-relativistic, magneto-hydrodynamic (GR-MHD) codes \citep[see, e.g.,][for a very partial list]{Igu+08, Narayan+12, Tchekho+12,  Liska+20, Porth+21, dihingia+23, GQZ+23, Lizhong+25, Fragile+25}. 
Yet, most of the currently existing GRMHD simulations are focused on ideal, non-radiative flows, which are appropriate for extremely low-luminosity sources. It is only recently that simulations including radiation 
% and non-ideal effects
have been conducted \citep[see, e.g.,][]{mckinney+14, sadowski+17, Liska+22, Chael+25, Wallace+25, Lizhong+25}, but their scope today is still limited, as inclusion of radiation both introduces computational complexity, and substantially increases the dimensionality of the parameter space. 

As of today, only a handful of works considered the role of radiative cooling \citep[e.g.,][]{Dibi+12, Yoon+20, dihingia+23, Salas+24, Singh25}.  This is despite the fact that cooling is an essential component that must be accounted for in describing realistic gas dynamics.
Although cooling is a mere loss of thermal energy, it fundamentally alters the disk's thermodynamics by reducing the gas pressure support. This, in turn, affects the disk dynamical evolution as it leads to reduction in the disk thickness, as well as changes the efficiency of jets and winds production and their properties.
Moreover, by changing the dynamics and morphology of accretion disks, radiative cooling imprints itself on the observable emission properties of the flow. 

These effects have been explored in the literature by several authors. For example, \citet{Dibi+12}, in the context of Sagitarius A*, found that even for a mass accretion rate as small as $10^{-7}\dot M_{\rm Edd}$, cooling affects the dynamics. Here, $\dot M_{\rm Edd}$ is the Eddington accretion rate.  Later, 3D simulations by \citet{Yoon+20}
confirmed its impact on disk structure and angular momentum transport. They showed that the overall flux, including the peak values at the
sub-mm and the far-UV, is slightly lower as a consequence of a decrease in the electron temperature. \citet{dihingia+23} highlighted the
importance of two-temperature treatments in modeling electron–ion thermodynamics in both SANE and MAD disks.
In that work, they observed that an increase of the accretion rate leads to electrons that are colder near the BH
but hotter far from it. 
Later, \citet{Salas+24} showed that synchrotron emission reduces the electron temperature in the inner disk and leads to a measurable ($\sim 10\%$) suppression of the (sub)mm synchrotron flux and its temporal fluctuations. 

In our previous study \citep{Singh25}, we presented an analytical framework and complementary GRMHD simulations to
demonstrate that in MADs, synchrotron cooling becomes dynamically significant above a mass accretion
rate $\sim10^{-5.5}\dot{M}_{\rm Edd}$. We showed that at higher accretion rates, 
the cooling rate due to synchrotron emission is sufficient to counterbalance the energy input from gravitational infall,
leading to a notable change in the disk structure, a modulation of the MAD parameter, $\phi_B$ (ratio of magnetic flux to the mass accretion rate), and the jet efficiency,
$\eta_{\rm jet}$ (ratio of energy flux to mass accretion rate). 
These results underscored the role of radiative losses in reshaping the force balance within the disk:
as cooling becomes more effective, the thermal pressure support diminishes, and the magnetic stresses and the dynamical
force become increasingly dominant in sustaining the disk equilibrium.

Here, we further study the effect of cooling on the internal structure of accretion disks in MAD state. In the inner disk regions of a MAD, the gas inflow becomes filamented. As we show here, the structure of these filaments strongly depends on the radiative properties, which, in turn, depend on the mass accretion rate. 
We demonstrate that the traditional definition of disk height can be misleading for MADs. We propose an alternative
definition based on the position of the density maximum, which more accurately captures the disk behavior under the influence of radiative
cooling. We find that the radiative efficiency, and the inner disk scale heights show a transition above a critical mass accretion rate. At low accretion rate, the radiative efficiency remains low, and the disk height is unaffected by the cooling, while at higher values, the disk height sharply decreases as the cold filaments are compressed by the external magnetic pressure.  To isolate these effects, in the current work we limit the discussion to disks around non rotating BHs, namely with spin parameter $a = 0$. 

The letter is structured as follows. Section \ref{sec:numerics} provides a review of the
governing GRMHD equations and their numerical implementation in cuHARM. In this section, we also present the
prescription for synchrotron and bremsstrahlung cooling and the initial conditions we use. Section \ref{sec:Results}
presents the results of three-dimensional (3D) GRMHD simulations. In section \ref{sec:cooling-derivation},
we derive the conditions under which cooling becomes important, and leads to substantial changes in the dynamics of accretion disks.
Finally, in Section \ref{sec:discussion}, we describe our conclusions and discuss future prospects.

\section{Numerical Methods}
\label{sec:numerics}
\subsection{Numerical Setup}

The numerical simulations presented in this work are carried out using the GPU-accelerated
code cuHARM \citep{Begue2023}. Similarly to HARM \citep{Gammie_2003, Noble+06},
cuHARM uses the finite volume method to numerically evolve the GRMHD equations
\citep[for reviews, see e.g.][]{marti+03, Rezzolla+13, Marti+15}. 
Under the ideal-MHD assumption, these are the conservation of mass, energy and momentum, as well as the homogeneous
Maxwell’s equations, respectively written as \citep{Gammie_2003}:
\begin{align}
    & \partial_t (\sqrt{-g}\rho u^t) = - \partial_i (\sqrt{-g}\rho u^i),\\
     &\partial_t (\sqrt{-g}  T^{t}_{\phantom{t}\nu}) =  - \partial_i (\sqrt{-g}  T^{i}_{\phantom{t}\nu}) + \sqrt{-g}  T^{\kappa}_{\phantom{\kappa}\lambda} \Gamma^{\lambda}_{\nu \kappa} - \sqrt{-g}u^{0}u_{\nu}\Lambda, \label{eq:EandPconserv}\\
      &  \partial_{t}(\sqrt{-g} B^i) = -\partial_j(\sqrt{-g} (b^ju^i - b^iu^j) \label{eq:induction},
\end{align} 
Here, $i$\footnote{Greek indices span over space and time (0, 1, 2, 3), while Roman indices pertain to space alone.} represents the spatial coordinates while $t$ is the time.
In these equations, $\rho$ is the gas density, $u^t$ and $u^i$ are the time and spatial components of the four velocity, 
$T^{\mu \nu}$ is the stress-energy tensor and $g$ is the determinant of the metric tensor $g_{\mu \nu}$. The magnetic field
is  $B^i = {}^\star F^{it}$, while $b^\nu \equiv u_\mu {}^\star F^{\mu \nu}$ denotes the magnetic field four vector,
where ${}^\star F^{\mu \nu}$ is the dual of the Faraday tensor. 
The last term on the right-hand side of Equation \ref{eq:EandPconserv} is the cooling term, expressing the  modification of the energy and momentum conservation equation due to radiative cooling losses. In this case, the cooling rate $\Lambda = \Lambda(\rho,~T_e,~b^2)$ depends on the gas density $\rho$, its temperature $T$,\footnote{More specifically on the electron temperature $T_e$.} and on the local magnetic field energy density,  $b^2$ (see further details in Section \ref{sec:cooling} below).

The stress-energy tensor $T^{\mu}_{\phantom{t} \nu}$ has the form  
\begin{equation}
\label{eq:energy-momrntum}
    T^{\mu}_{\phantom{t} \nu} = (h + b^2)u^{\mu}u_{\nu} + \left(p + \dfrac{1}{2}b^2\right)\delta^{\mu}_{\phantom{t} \nu} -b^{\mu} b_{\nu},
\end{equation}
where $h = (\rho + \epsilon + p) $ is the enthalpy, $\epsilon$ is the gas energy density, $p$ is the gas pressure, and
$b^2 = b_{\mu}b^{\mu} \equiv 2 p_b$ is the square of magnetic field four-vector, which equals two times the magnetic pressure $p_b$. 
Note that a factor of $\sqrt{4 \pi}$ is implicitly included in the electromagnetic field.

The no-monopole constraint on the magnetic field $\Vec{\nabla} \cdot \Vec{B} = 0$ is written in the form
\begin{align}
     \partial_{i}(\sqrt{-g}B^i) = 0,
\end{align}
and is numerically satisfied with the flux-CT method \citep{Toth+00,Gammie_2003}.

In the simulations presented in this letter, we adopt the Kerr-Schild (KS) coordinate system defined by ($t, r, \theta, \phi$). To enhance the accuracy of our calculations close to the BH and along the equator, we use the modified Kerr-Schild (MKS) coordinates ($t, q^1, q^2, \phi$) including cylindrification close to the pole at small radii \citep{Mckinney+04, Tchekho+11}. 
The computational grid for the numerical simulation is defined with inner and outer disk boundaries at
$r_{\min} = 0.87r_{H}$, $r_{\max} = 5 \times 10^3 r_{g}$ respectively.  Here, $r_{H} = 2 r_g \equiv 2 GM/c^2$ is the event horizon radius of a non-spinning BH with $M$ being its mass. 
In addition, the coordinates $\theta$ and $\phi$ are such that $\theta \in [ 0, \pi]$, and
$\phi \in [ 0, 2\pi]$. For the magnetohydrodynamics, we utilize a system of units where the speed
of light $c$, and the gravitational constant $G$, and the BH mass are normalized to 1, i.e., $G = c = M = 1$.
The full details of the numerical scheme is described in \citet{Begue2023}.

\subsection{Cooling processes}
\label{sec:cooling}

In this letter, we consider sub-Eddington mass accretion rate up to $\dot M \lesssim 10^{-4} \dot M_{\rm Edd}$. Therefore,
radiative cooling is calculated by considering the synchrotron
and bremsstrahlung emission processes. We ignore cooling by Compton scattering, as it was found to be
sub-dominant at all accretion rates considered in this paper \citep{Dibi+12, Yoon+20, dihingia+23},
and furthermore requires knowledge of the photon field, which we cannot calculate with the current code setup;
see however \citet{Wallace+25} for full radiation transport solution carried with cuHARM, as well as \citet{asahina+20,white+23}.
In this subsection we use real physical units.

We define
\begin{equation}
    \Lambda = q^{-}_{br} +  q^{-}_{s}, 
\end{equation}
where $q^{-}_{br}$ and $ q^{-}_{s}$ are the cooling rates due to bremsstrahlung and synchrotron, respectively. 
We use the cooling prescription provided by
\citet{Esin+96} and \citet{2009ApJ...693..771F}. For the Bremsstrahlung process, we consider the contributions
from ion-electron and electron-electron interactions, 
\begin{equation}
    q^{-}_{br} = q^{-}_{ei} +  q^{-}_{ee},
\end{equation}
where
\begin{equation}
\label{First_equation}
  q^{-}_{ei}=\begin{cases}
    1.48 \times 10^{-22}n_p n_e \times  \left [ 4\sqrt{ \frac{2 \Theta_e}{ \pi^3}} \left ( 1 + 1.781\Theta_e^{1.34} \right ) \right ] \mathrm{erg\,cm^{-3}\,s^{-1}}, & \text{if $\Theta_e<1$}\\
    1.48 \times 10^{-22}n_p n_e \left (  \frac{9 \Theta_e}{2 \pi} [\ln(1.123\Theta_e + 0.48)+ 1.5] \right) \mathrm{erg\,cm^{-3}\,s^{-1}}, & \text{if $\Theta_e\gtrsim1$},
  \end{cases}
\end{equation}
and
\begin{equation}
\label{Second_eq}
    q^{-}_{ee} = \begin{cases}
    2.56 \times 10^{-22}n^{2}_e \Theta_e^{1.5}\left (1 + 1.1\Theta_e + \Theta_e^2 - 1.25\Theta_e^{2.5} \right) \mathrm{erg\,cm^{-3}\,s^{-1}}, & \text{if $\Theta_e<1$}\\
    3.42 \times 10^{-22}n^{2}_e \Theta_e \left [\ln(1.123\Theta_e) + 1.28\right ]  \mathrm{erg\,cm^{-3}\,s^{-1}}, & \text{if $\Theta_e\gtrsim1$}.
    \end{cases}
\end{equation}
In these equations, $\Theta_e = k_BT_e/m_ec^2$ is the dimensionless electron temperature where $T_e$ is the
electron temperature, $k_B$ is the Boltzmann constant, $m_e$ is the mass of the electron, and the $c$ is
the speed of light. In the calculations of the cooling rate, the protons thermal motion is neglected
\citep{Stapney+83}. In addition, we neglect pair creation, and therefore the density of electrons and
protons in the disk are equal, $n= n_p = n_e= \rho / (m_p+ m_e)$ (where $m_p$ is the proton mass) and is directly obtained from the density $\rho$ evolved in cuHARM.  At low mass accretion rates, $\dot{M} \lesssim10^{-4}\dot{M}_{\rm Edd}$, pair production is expected to be inefficient with the pair fraction remaining small \citep[$\lesssim 1\%, $][]{Chan+26}.

In the presence of a strong magnetic field, as relevant in MADs, synchrotron cooling is an
efficient mechanism for charged particles to radiate their energy. We adopt the optically thin synchrotron cooling rate and neglect synchrotron self-absorption (SSA). This approximation is appropriate for the current work, since the cooling function is used only as a local energy-loss term in the GRMHD evolution rather than for predicting the emergent radiation spectrum. While SSA can modify the low-frequency spectrum, it is not expected to significantly affect the integrated local cooling rate except in optically thick regions, which are not expected to dominate the cooling in our models.
The cooling rate by synchrotron
radiation (per unit volume) is given by \citep[e.g.][]{1986rpa..book.....R}
\begin{equation}
\label{sync_equation}
    q^{-}_s = \dfrac{4}{3} c \sigma_T \gamma_e^2 \beta_e^2 U_B  n_e \,\,\, \mathrm{erg\,cm^{-3}\,s^{-1}},
\end{equation}
Here, $\sigma_T$ is the Thomson cross-section, $\gamma_e$ is the electron Lorentz factor
(corresponding to the normalized velocity $\beta_e$), and $U_B = b^2/2$ (note that in the units used here, $U_b = p_b$).

Both the synchrotron and bremsstrahlung cooling rates depend on the electron temperature ($T_e$), which we
set as a constant fraction of the proton temperature; this assumption corresponds to a strong coupling between
electrons and protons.
We define  the model parameter $\tau = T_p/T_e$.
Previous works \citep{Dibi+12, Drappaeu+13, Yoon+20} used values of $\tau = 3 - 10$. In this work, we
set $\tau = 3$. 
The gas pressure, $p$ is related to the internal energy density by the ideal equation of state $p = (\Gamma - 1)u$, where $\Gamma$ the adiabatic index
of the plasma and is taken as $\Gamma = 14/9$ in all our simulations. The value of $\Gamma = 14/9$ is chosen to approximate the effective adiabatic index of a trans-relativistic electron-ion plasma. Using $p = p_e + p_p= n k_B (T_e + T_p)$, one finds 
\begin{equation}
        T_e = \dfrac{(\Gamma - 1)u}{\rho}\left(\dfrac{m_p + m_e}{k_B}\right)\left(\dfrac{1}{1 + \tau}\right).
\end{equation}
While the code uses normalized units in calculating the dynamics, inclusion of radiative cooling necessitates the
conversion to physical units. We follow \citet{2018A&A...613A...2B} and define a length unit $\mathcal{L} \equiv r_g = GM/c^2$
and a time unit $\mathcal{T} \equiv r_g/c = GM/c^3$, which only depend on the mass $M$ of the BH. We further define a mass unit,
which depends on a pre-assumed mass accretion rate $\mathcal{M} = \kappa \dot{M}_{Edd}\mathcal{T}$. In these definition, $\kappa$ is the assumed accretion rate, $\dot{M}_{Edd}$ is the Eddington accretion rate given by $\dot{M}_{Edd} = L_{Edd}/\eta c^2 = 1.33 \times 10^{18}~(M/M_{\odot})~{\rm g~ s^{-1}}$ where $L_{\rm Edd}$ is the Eddington luminosity, $M_{\odot}$ is the solar mass and $\eta$ is the radiative efficiency, which is taken here to be 10\%.
We then define two conversion factors between code unit and physical units $\bar \rho_{ph} = \mathcal{M/L}^3$ 
such that the physical unit density is $\rho_{\rm ph} = \bar \rho_{ph} \rho$, $n_{e,ph} = \rho_{ph} / (m_p + m_e)$, 
and the energy density unit is 
$u_{ph} = \bar \rho_{ph}c^2$. In the definitions above, the subscript "$ph$" refers to a quantity with physical units (in cgs).

\subsection{Initial conditions}
\label{sec:physical_setup}

We begin our simulations by initializing an un-magnetized axisymmetric torus in hydrostatic equilibrium 
around a non-rotating BH following \citet{Fishbone+76}. The torus is defined by two parameters, $r_{\rm in}$
and $r_{p,\max}$, representing the inner radius and the radius of the pressure maximum of the torus. In this work, we
consider large disks with $r_{\rm in} = 20 r_g$ and $r_{p,\max} = 41 r_g$. The basic disk model is
rendered unstable in order to initiate accretion through the development of magneto-rotational
instability \citep{Balbus+98} by introducing minor random perturbation of magnitude 4\% to the gas pressure, similar to \citet{Porth+19}. 
In order to set up the magnetic field in the system, we used in all cases a single-loop
magnetic field specified by the vector potential
\begin{equation}
    A_{r} = A_{\theta} = 0, \\ 
    \hspace{0.2cm} \text{and} \hspace{0.2cm} A_{\phi} = \text{max}\left\{0, \left(\dfrac{\rho}{\rho_{max}}\right) \left(\dfrac{r}{r_{in}} \text{sin} \theta\right)^3 \text{exp} \left(-\dfrac{r}{400}\right) - 0.2\right\}.
\end{equation}
This initial magnetic field configuration has been previously applied by, e.g., \citet{Wong+21,Narayan+22,GQZ+23}. We adopt it
here, since this specific parameter combination is known to lead to a MAD accretion flow. The normalization of
the initial magnetic field is chosen such that the initial ratio of maximum gas pressure to maximum magnetic pressure is 100.

In this work, we used the series of simulations with spin $a = 0$
presented in \citet{Singh25}. These simulations
target different physical mass accretion rates between $10^{-7} \dot M_{\rm Edd}$
and $10^{-4} \dot M_{\rm Edd}$.
As explained in section \ref{sec:cooling} above, when radiative cooling is
included in the simulation, two physical scales
need to be defined to calculate the cooling rate. The first is the mass
of the central BH, which sets the characteristic length and time scales of system, as well as the Eddington
accretion rate. We choose the BH mass to match that of Sgr A*, using $ M_{\rm BH} = 4 \times 10^6~M_\odot$ \citep{EHT+22a}. 
The second required scale is a reference density used to determine the gas density. This is necessary to convert the
simulated accretion rate from code units to physical units. To determine this scale, we first run a simulation
without cooling to measure a baseline accretion rate. This result is then scaled to match a desired physical
accretion rate, expressed as a fraction of the Eddington rate $ \dot{M}_{\rm Edd} $. We then perform the simulations with radiative cooling included, using the determined scaling factor.

The details of the 5 simulations used in this work are given in Table \ref{Simulation-Details}.
The simulations are conducted until a final time of $3 \times 10^4 \, t_g $,
where $ t_g = r_g/c $. This is long enough for the MAD
regime to be fully established and for a steady-state balance between inflow and outflow to be attained,
up to radii of a few tens of $r_g$. For all the results presented below, we compute the time averages of the diagnostics over
the interval from $ t = 1.5 \times 10^4 \, t_g $ to $3 \times 10^4 \, t_g $.

\begin{table}
\begin{center}
\begin{tabular}{|c|c|c|c|c|}
\hline
Name  & Cooling & Target $\dot{M}/\dot{M}_{\rm Edd}$ & $\langle \dot{M}/\dot{M}_{\rm Edd} \rangle$ & $\langle \phi_B \rangle$ \\
\hhline{|=|=|=|=|=|}
S-4 & On  & $10^{-4}$ & $1.48 \pm 0.79 \times 10^{-4}$ & $19.78 \pm 7.83$ \\
S-5 & On  & $10^{-5}$ & $0.97 \pm 0.59 \times 10^{-5}$ & $26.02 \pm 5.69$ \\
S-6 & On  & $10^{-6}$ & $0.90 \pm 0.58 \times 10^{-6}$ & $30.39 \pm 5.41$ \\
S-7 & On  & $10^{-7}$ & $0.84 \pm 0.55 \times 10^{-7}$ & $31.78 \pm 7.11$ \\
SNC & Off & --        & --                            & $33.02 \pm 8.02$ \\
\hhline{|=|=|=|=|=|}
\end{tabular}
    \caption{ List of simulations along with their initial parameters. Each simulation assumes a non-spinning BH of mass $4 \times 10^{6}\, M_{\odot}$, and an initial plasma-$\beta$ parameter of $100$. Each simulation has a resolution of $256 \times 128 \times 256$ along $N_r \times N_{\theta} \times N_{\phi}$. The last two columns give the time average values of accretion rate $\dot{M}$, and the MAD parameter $\phi_{B}$. This average is calculated for times such that $1.5 \times 10^4 t_g < t < 3.0 \times 10^4 t_g$.}
    \label{Simulation-Details}
    \end{center}
\end{table}

\section{Results}
\label{sec:Results}

\begin{figure}
    \centering
    \includegraphics[width=0.99\linewidth]{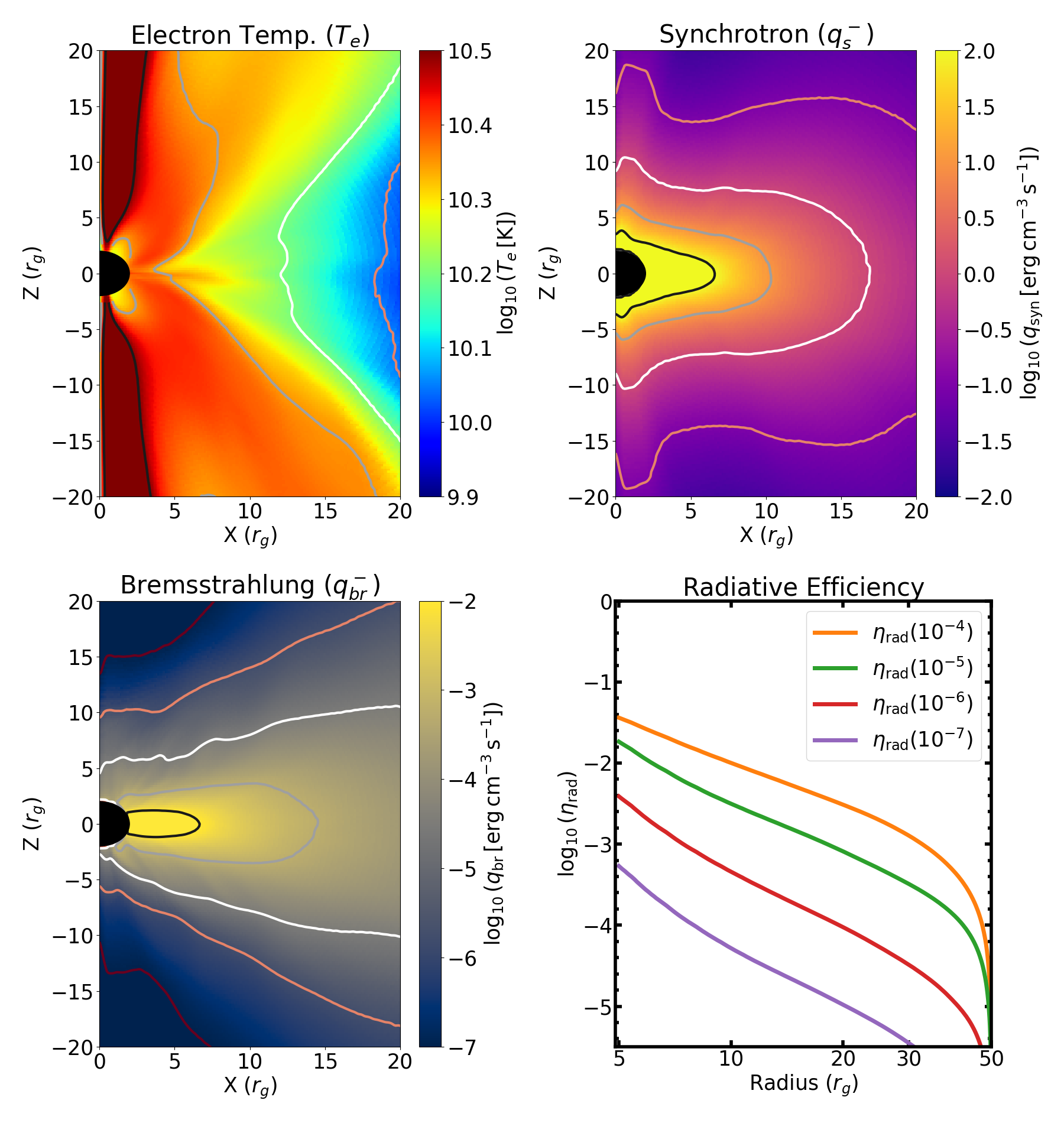}
    \caption{Time-average maps of the electron temperature (upper-left), comoving synchrotron (upper-right) and
    bremsstrahlung power (lower-left) for the simulation S-4 with accretion rate $10^{-4} \dot{M}_{ \rm Edd}$. The averages shown in all panels are calculated over the time interval $1.5 \times 10^4 \mathbf{t_g} < t < 3.0 \times 10^4 \mathbf{t_g}$. We note
    the different scaling for the synchrotron and bremsstrahlung processes.
    It is therefore  observed that synchrotron cooling dominates the bremsstrahlung cooling by several orders
    of magnitude. The electron temperature is low in the disk midplane. In contrast, the
    temperature in the region close to pole is very high which is associated to the numerical flooring. A
    similar behavior is observed across all mass accretion rates. \textit{lower-right} panel shows the radiative efficiency $\eta_{\rm rad}(r)$ as a function of radius for different mass accretion rates $\dot{M}$ for all the simulations. The radiative efficiency increases with the increase in the mass accretion rate due to enhanced cooling. Conversely, it decreases with increasing radius, where the disk becomes less efficient at radiating energy.}
    \label{Cooling components}
\end{figure}

\subsection{Dominant cooling process}
We present in Figure \ref{Cooling components} (first three panels) time-averaged polar maps of the
electron temperature, synchrotron power, and bremsstrahlung power for the model S-4 with
$\dot{M} = 10^{-4}\dot{M}_{\rm Edd}$, corresponding to the largest mass accretion rate
considered in this letter. The electron temperature (left panel) shows a clear vertical stratification:
it is low in the dense disk body near the equatorial plane and decreases gradually with radius. At higher
latitudes, the temperature increases toward the wind region, where no numerical floors are applied.
However, the very high temperatures seen along the jet axis are not physical as resulting
from the imposed density and internal-energy floors in the funnel. 

The top-right and bottom-left panels in Figure \ref{Cooling components} compare the synchrotron and bremsstrahlung power. Synchrotron cooling clearly dominates the radiative losses throughout the domain (note the different color scales). 
It peaks in the inner disk near the black hole and along the equatorial plane, then declines rapidly with radius. This behavior follows from the strong dependence of the cooling rate on the local electron temperature, magnetic field strength, and density, all of which are largest in the inner disk and have a strong radial dependence. In contrast, bremsstrahlung cooling is much weaker everywhere and exhibits a more spatially extended distribution, with relatively higher values toward the dense midplane due to its sole density and temperature  dependence. Across all accretion rates considered in this work, synchrotron emission remains the dominant cooling mechanism. Similar conclusion also appeared in recent studies such as \citet{Liska+24,Salas+24}. 

\subsection{Radiative Efficiency}
\label{sec:rad_eff}

To illustrate the importance of radiative losses on the dynamics of the accretion flow, we calculate the radiative
efficiency $\eta_{\rm rad}$ in the equatorial region, as particles
accrete from 50$r_g$ inward to some radius $r$. It is written as:
\begin{align}
     \eta_{rad}(r) = \dfrac{2\pi \displaystyle  \int_{r^{'} = r}^{50r_g} \displaystyle  \int_{4\pi/9}^{5\pi/9} \sqrt{-g} d\theta dr^{'}  \langle q^- \rangle_{t,\phi}} {\langle \dot{M}_{5r_g}\rangle}. \label{eq:rad_eff}
\end{align}
Here, $\langle q^- \rangle_{t,\phi}$ is the time and azimuthal average of the radiated energy
per unit volume and $\langle \dot{M}_{5r_g}\rangle$ represents the time averaged mass accretion rate measured at radius of $5r_g$. It is measured as
$\dot{M}_{5r_g} = - \int_{0}^{\pi}\int_{0}^{2\pi}
\left.\rho u^r\right|_{r=5r_g}
\,\sqrt{-g}\,
\mathrm{d}\theta\,\mathrm{d}\phi$.
The radius $r=5r_g$ is chosen to minimize contamination from the numerical density floor.
The value of $50 r_g$ for the outer boundary is chosen as the radiative efficiency becomes negligible beyond this radius
and since it roughly corresponds to the inflow-outflow equilibrium radius. The integration boundaries here
are reversed relative to similar calculations carried in previous works \citep[e.g.,][]{Liska+23, Salas+24}. This is to
reduce the sensitivity of the radiative efficiency $\eta_{\rm rad}$ to the choice of the integration boundary as
the largest contributions originates from small radii. Finally, note that we carried out this calculation down to
$r = 5 r_g$, and not below, in order to filter out contributions from the density floor. The angular domain $\theta \in [4\pi/9,\, 5\pi/9]$ is chosen to isolate the dense disk component while excluding low-density polar wind regions that dominate the volume but contribute negligibly to mass accretion and radiative output in this setup. To assess sensitivity, we have recomputed $\eta_{\mathrm{rad}}$ using wider angular wedges up to $\pm 30^\circ$. We find that $\eta_{\mathrm{rad}}$ varies by less than $\sim 10\%$, indicating that the radiative efficiency is robust.

The bottom-right panel of Figure \ref{Cooling components} shows the radiative efficiencies $\eta_{\rm rad}$
for all simulations (namely mass accretion rates) in this work. As expected, the radiative
efficiency increases with increasing mass accretion rate. We observe that there is a non-linear increase in the radiative efficiency with mass accretion rate. At low $\dot{M}$ it increases linearly with $\dot M$ at almost all radii before it starts to saturate near $10^{-4} \dot{M}_{\rm Edd}$.
It is also observed that the radiative efficiency is higher in the region closer to the BH. 
This is consistent with the fact that most of the radiative losses occur in the
innermost regions of the disk where (i) the gravitational energy release is the highest,
while (ii) plasma temperature and magnetic fields are high, enhancing
the synchrotron cooling rate.
At a given mass accretion rate,
$\eta_{\rm rad}$ exponentially increases towards the BH.
However, at low accretion rates, e.g. $10^{-7} \dot{M}_{\rm Edd}$, the radiative efficiency remains negligible at all radii, leading to a picture consistent with radiatively inefficient accretion flows (RIAFs), where most of the thermal energy is accreted by the BH rather than radiated away. In contrast, at higher mass accretion rate, the disk becomes radiatively efficient. We find that the radiative efficiency is around $\approx$3.6\% at $r = 5r_g$ for S-4.
Most importantly our results offer new insights into the regimes of small ($\sim 10^{-7} \dot M_{\rm Edd}$) to intermediate ($\sim 10^{-4} \dot M_{\rm Edd}$) mass accretion rate that have remained relatively unexplored, with previous studies focusing on small mass accretion rates \citep[{see e.g.}][]{Liska+24, Chael+25}.

\subsection{Disk scale height}
\label{sec:height}

A commonly adopted diagnostic for the vertical thickness of accretion disks is the density-weighted angular scale height \citep[see e.g.][]{Porth+19,Narayan+22, Chael+25},
\begin{equation}
\label{eq:old-disk-height}
\frac{h}{r}(r) = 
\frac{\displaystyle \int_{t_{\rm beg}}^{t_{\rm end }} \int_{0}^{2\pi} \int_{0}^{\pi} \left| \frac{\pi}{2} - \theta \right| \rho (r, \theta, \phi, t)  \sqrt{-g}~d\theta~d\phi~dt}
{\displaystyle \int_{t_{\rm beg}}^{t_{\rm end }} \int_{0}^{2\pi} \int_{0}^{\pi}  \rho (r, \theta, \phi, t) \sqrt{-g}~d\theta~d\phi~dt}.
\end{equation}
It measures the mean angular deviation of the mass distribution from the equatorial plane.
This definition is well motivated and widely used in simulations of weakly magnetized (SANE)
disks, where the density distribution is approximately symmetric about the equator. 
However, this diagnostic implicitly assumes that the equatorial plane coincides with the
physical midplane of the disk. This condition, though, is not generally satisfied in MADs.

\begin{figure}
    \centering
    \includegraphics[width=0.8\linewidth]{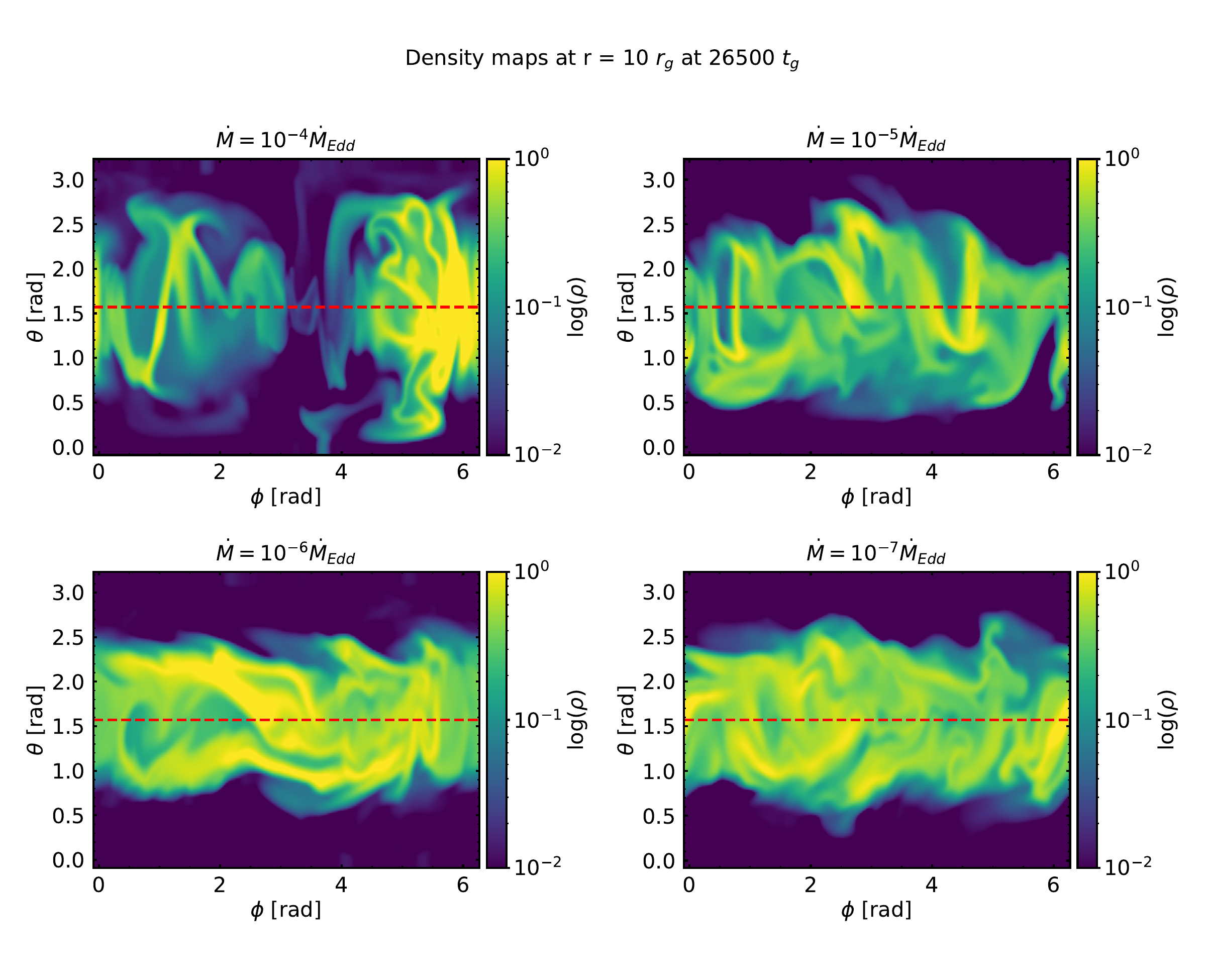}
    \caption{ $\theta-\phi$ maps of the density (in code units) at $r = 10 r_g$ for simulations with
    decreasing mass accretion rates, ranging from $10^{-4}~\dot{M}_{\rm Edd}$ (top left) to $10^{-7}~\dot{M}_{\rm Edd}$ (bottom right),
    shown at time $t = 2.65 \times 10^4 \mathbf{t_g}$. The dashed red line marks the equatorial plane. At all azimuth, the density maxima
    are clearly offset from the midplane. Strong fluctuations in density are observed as the mass accretion rate increases,
    indicating that the disk becomes increasingly turbulent and filamentary as radiative cooling increases.}
    \label{fig:funnels}
\end{figure}

Rather, in MADs, the accumulation of strong vertical magnetic
flux near the BH leads to a magnetically dominated equatorial region and drives interchange
instabilities that produce intermittent, filamentary accretion streams \citep{Narayan+12}.
This is shown in Figure \ref{fig:funnels} which presents $\theta$–$\phi$ maps of the 
gas density in code units at $t = 2.65 \times 10^4~t_g$ for simulations S-4 to S-7 with mass accretion rates ranging from $10^{-7}\dot{M}_{\mathrm{Edd}}$ to $10^{-4}\dot{M}_{\mathrm{Edd}}$. For this figure, the radius is set to $10~r_g$. In each panel, the dashed red line indicates the equatorial plane. 
As seen from the figure, the accretion filaments are not confined to the equatorial plane, breaking the equatorial symmetry in the inner disk regions.
As a result, the gas density distribution does not peak at $\theta = \pi/2$, but rather fluctuates  above and below the equator. 
At the lowest accretion rate ($10^{-7}\dot{M}_{\mathrm{Edd}}$), the density distribution appears relatively smooth and vertically extended, with only mild azimuthal variations.

As the accretion rate increases, the radiative cooling leads to compression of the accreting gas, resulting in sharper density variations
and a more fragmented appearance. Indeed, 
we observe that simulations with large mass accretion rate produce a mildly larger magnetic field (in code units) close to the BH than small mass accretion rate simulations. This effect, though, is not sufficient to explain the large variations of plasma $\beta = p_{\rm gas}/p_{\rm mag}$. 
In fact, at high accretion rates, cooling becomes  significant enough to cause gas pressure loss, reflected by a decrease in the plasma-$\beta$ parameter. 

For S-4, the simulation with the hightest mass accretion rate, the density map is dominated
by strong, irregular fluctuations. The flow exhibits highly turbulent, filamentary structures with substantial
displacement of the densest region away from the equator.
Overall, Figure \ref{fig:funnels} illustrates a clear trend: as
the mass accretion rate increases, the disk transitions from a relatively smooth, vertically extended configuration
to a more turbulent and un-structured state, at $r = 10 r_g$.

In a filamented flow, the density-weighted average given in Equation \ref{eq:old-disk-height}
does not describe the intrinsic geometric thickness of the accretion flow. Instead, it reflects the displacement of dense filaments away from the equatorial plane and their average vertical position. As a result, variations in the vertical location of these filaments produce variations in the inferred disk height ($h/r$), even if the actual thickness of the individual streams does not change significantly. Therefore, the standard definition does not provide a reliable measure of the disk thickness in MAD simulations. This also makes direct comparisons between SANE and MAD flows difficult: as there is no well-defined single midplane, a disk scale height defined with respect to the equatorial plane becomes ambiguous.

For this reason, we introduce an alternative diagnostic that captures the physical size of the filaments and does not assume either equatorial symmetry or a density maximum at $\theta = \pi/2$. 
In this alternative definition, we use the following algorithm. First, at each radius $r$ and time $t$, we identify the maximum density $\rho_{\rm \max, global}$ throughout the $\theta$-$\phi$ domain. The cells with densities below a fraction $f_{\mathrm{floor}} = 10$ of this maximum,
$\rho < \rho_{\rm \max, global} / f_{\mathrm{floor}},$
are excluded from further analysis. This threshold is used to identify contiguous regions of significant density throughout the domain.
consider each azimuthal slices independently.
For each given $\phi$, we analyse the density $\rho(\theta; r, \phi)$ on the $(\theta)$ ring ($0 \leq \theta \leq \pi$). The cells on this ring that satisfy the density threshold condition defined above ( $\rho \geq \rho_{\rm \max, global}/f_{\mathrm{floor}}$) are grouped into contiguous regions (which we call islands) in $\theta$. 
Within each island, the location of the local density maximum is determined, and the corresponding polar angle $\theta_{\max, \rm{local}}$ is recorded. Only islands whose peak density exceeds $\rho_{\max}/f_{\mathrm{peak}},$
with $f_{\mathrm{peak}} = 7$, are retained. The remaining islands are discarded, as we assume they correspond to weak fluctuations or low density structures rather than the main disk body.

For each retained island, we compute a local thickness in the $\theta$-direction,
\begin{equation}
\label{eq:new-height}
\left(\frac{H}{r}\right)_{\mathrm{island}}(r,\phi,t)
=
\frac{\displaystyle \int_{\theta_1}^{\theta_2}\rho \sqrt{-g}\,
\left| \theta - \theta_{\max,\mathrm{local}} \right| \, d\theta } 
{\displaystyle \int_{\theta_1}^{\theta_2} \rho_{\rm max, global} \sqrt{-g}\, d\theta}.
\end{equation}
Here, $\theta_1$, $\theta_2$ denote the limits of the polar angle for a given island, and the integrals are taken over the $\theta$ extent of the island. When multiple qualifying islands are present at the same azimuth, their thicknesses are averaged to obtain a single height. %$H_{\phi}(r)$. 
The disk height at radius $r$ and time $t$ is then defined as the azimuthal average, $H(r,t) = \left\langle  H_{\mathrm{island}}(r,\phi, t) \right\rangle_{\phi}$,
and is subsequently time-averaged over the desired interval.
We verified that moderate variations in the values of the two parameters used in our definition of the disk height, namely $f_{\mathrm{floor}}$ and $f_{\mathrm{peak}}$ 
do not quantitatively affect the resulting height profiles as long as the values are kept between 5-15. 

This method better captures the geometrical structure of the inner accretion flows. 
By identifying coherent density structures on a per-azimuth basis, the method avoids artificial thickening when multiple off-equatorial filaments are averaged together. Low-density coronal regions and magnetically dominated cavities are excluded by construction, because the analysis involves density-weighted averaging, which suppresses contributions from regions of low plasma density. So the measured height reflects the geometry of the accreting matter itself.  In the outer regions, where the density distribution is expected to peak near the equator, we expect the thickness measured by our newly introduced method to approach values similar to those obtained with the standard density-weighted measure, which it does. The disk height obtained using our new definition in this region is about a factor two smaller than the old definition.
This is because of the global normalization we use ($\rho_{\rm max, global}$) which concentrates the disk around its center of mass.

\begin{figure}
    \centering
    \includegraphics[width=0.9\linewidth]{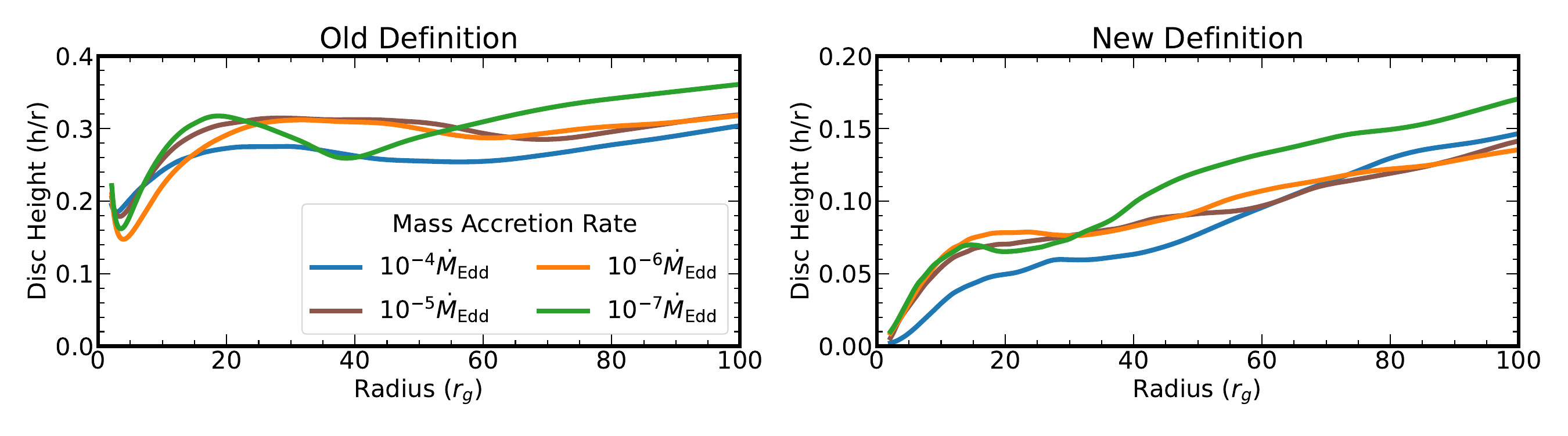}
    \caption{Disk scale height ($h/r$) as a function of radius using the standard definition (left) and the revised definition (right). The standard diagnostic suggests that the scale height is largely unaffected by radiative cooling. In contrast, the revised definition clearly shows a significant decrease in disk thickness once the accretion rate reaches $\dot{M}_{\rm crit}$, capturing the physical compression of the flow that the traditional measure obscures.}
    \label{fig:disk-height}
\end{figure}

The difference between the two definitions is demonstrated in Figure \ref{fig:disk-height}, where we examine the disk height $h/r$ averaged over times $2 \times 10^4 - 3 \times 10^4~ t_g$ obtained using the traditional density-weighted angular disk height (equation \ref{eq:old-disk-height}-left panel) and the maximum-density-based disk height  (equation \ref{eq:new-height}-right panel). 
The left panel of Figure \ref{fig:disk-height} shows the radial profile of the disk height over a range of mass accretion rates, based on a density-weighted measurement. It can be observed that at large radii, the disk height remains nearly constant across all accretion rates, down to a radius of approximately $15r_g$, below which it decreases sharply by about 50\%. Notably, the height does not approach zero close to the BH. An apparent increase in disk thickness near $5r_g$ is also visible; this feature, seen in various works \citep[e.g.,][]{Narayan+22,GQZ+23,Chael+25} is an artifact of the measurement, arising from filamentary structures within the disk and their displacement away from the equator.

In the right panel of Figure \ref{fig:disk-height}, we show the disk height using our modified definition.
We find that the disk height decreases monotonically with radius towards the BH.  Importantly, the disk height
approaches zero as the radius decreases toward the BH, a trend not captured by the old definition. We verified that this decrease in disk height is not solely due to the $\rho_{\max}$ normalization, as the same qualitative trend is recovered with an alternative, unnormalized definition. This indicates genuine physical thinning of the plunging flow, although its precise magnitude depends on the adopted height definition.
In the outer region ($r \gtrsim 40r_g$), the disks of all simulations have comparable thickness,
showing that cooling does not impact the dynamics at these distances.  We point out that at these radii,
no significant disks filamentation exist for any of the simulations. At radii $r \lesssim 40 r_g$, the disk
of S-4 becomes markedly thinner compared to that of S-5, S-6 and S-7, demonstrating the role of radiative cooling up to
this radius for $\dot M = 10^{-4} \dot M_{\rm Edd}$. 
Instead, the disks thickness of S-5, S-6 and S-7 remain comparable at all radii demonstrating that they do not efficiently cool. The larger the mass accretion rates, the larger the radius up to which cooling reduces the disk height. 

A major difference arises in the inner region, where magnetic flux accumulates and substantial disk filamentation takes place. In our simulations, we observe those filamentation in the inner $\sim 30-40~r_g$.
In this region, there is a considerable difference in disk height between disks with a very high accretion rate ($10^{-4}\dot{M}_{\rm Edd}$) compared to those with lower accretion rates.  
This behavior is attributed to enhanced radiative cooling up to these radii: the disk loses internal energy through radiation, which reduces thermal pressure support and leads to a substantial decrease in the scale height in the inner regions. This behavior is detailed in the following section. 

\section{Critical accretion rate above which cooling substantially affects the disk structure}
\label{sec:cooling-derivation}

The results presented in Figure \ref{fig:disk-height} show a substantial modification to the disk height at $\dot{M} = 10^{-4}\dot{M}_{\mathrm{Edd}}$, particularly prominent within the innermost $\sim 30-40\,r_g$. 
This occurs, as at high accretion rate, substantial radiative cooling causes considerable loss of gas pressure, enforcing strong filamentation at small radii, as seen in Figure \ref{fig:funnels} above. Here, we analytically derive the conditions under which radiative cooling significantly affects the disk structure.

In the inner regions of the disk, heating is assumed to arise from strong coupling between the gas and magnetic field, which is mediated by turbulence, shocks, and magnetic reconnection. Under this assumption, and adopting approximate equipartition, the specific internal energy of the gas can be estimated as $u_{\mathrm{gas}} \sim u_B/\rho$, where $u_B$ is the magnetic energy density.
The gas heating rate can therefore be approximated as:
\begin{align}
\frac{d u_{\text{gas}}}{dt} 
&= \frac{d}{dt} \left( \frac{u_B m_p}{\rho} \right)
= v_r \frac{d}{dr} \left( \frac{u_B m_p}{\rho} \right) 
\approx  \frac{u_B m_p v_r}{\rho r}.
\end{align}
where $v_r$ is the radial velocity of the flow.

We consider synchrotron radiation as the dominant cooling mechanism (see Section~\ref{sec:cooling}), with the power emitted by a single electron of Lorentz factor $\gamma_e$ (and corresponding velocity $\beta_e = v_e/c$) given by Equation~\ref{sync_equation}. 
Comparing the heating and cooling rates per particle, under the assumption of strong coupling between electrons and protons, leads to:
\begin{equation}
\frac{4}{3} c \sigma_T (\gamma_e^2 \beta_e^2) u_B = \frac{u_B m_p v_r}{\rho r}.
\label{eq:heat_cool}
\end{equation}

The electron's momentum can be expressed as follows. Assuming ultra-relativistic thermal electrons, the average electron momentum is proportional to the electron's temperature, 
$\langle \gamma_e^2 \beta_e^2 \rangle \approx \langle \gamma_e^2 \rangle = 12 \Theta_e^2
$, where the factor of 12 arises from considering the Maxwell-Jüttner distribution \footnote{We point out that in the calculation presented here, 
for simplicity we adopted the ultra-relativistic approximation $12\Theta_e^2$ for the electron momentum term. Substituting the exact relativistic expression—derived from the moments of the Maxwell-Jüttner distribution, results in a negligible correction of approximately $8-10\%$, well within the precision of the numerical model.}
.
Here we assume a one-temperature GRMHD model with electron temperature directly related to the proton temperature via $T_e = T_p / \tau$ (see section \ref{sec:cooling}), leading to 
\begin{equation}
\Theta_e = \frac{k_B T_e}{m_e c^2} = \frac{k_B (T_p / \tau)}{m_e c^2} = \frac{m_p}{\tau m_e} \left( \frac{k_B T_p}{m_p c^2} \right) = \left( \frac{m_p}{m_e} \right) \left( \frac{\Theta_p}{\tau}\right). 
\end{equation}

Substitute $\gamma_e^2 \beta_e^2 = 12 \left( \frac{m_p  \Theta_p }{\tau m_e}\right)^2$ into Equation \ref{eq:heat_cool}  and simplifying:
\begin{equation}
16 c \sigma_T \left( \frac{m_p \Theta_p}{\tau m_e} \right)^2 = \frac{m_p v_r}{\rho r}.
\label{eq:heating_cooling2}
\end{equation}
Expressing the density via the accretion rate $\rho = \dot M / (4 \pi r^2 v_r)$, 
and further using the definition of the Eddington accretion rate, 
$\dot{M}_{\text{Edd}} = (4 \pi G M m_p)/(\eta \sigma_T c)$ and the gravitational radius $r_g = GM/c^2$, Equation \ref{eq:heating_cooling2} takes the form 
\begin{equation}
\frac{\dot{M}}{\dot{M}_{\text{Edd}}} = \frac{\eta}{16} \left( \frac{r}{r_g} \right) \left( \frac{v_r}{c} \right)^2 \left( \frac{\tau}{\Theta_p} \right)^2 \left( \frac{m_e}{m_p} \right)^2.
\label{eq:final}
\end{equation}

Equation \eqref{eq:final} can be further simplified by expressing it using the Keplerian velocity $ v_K = \sqrt{GM/r} = c \sqrt{r_g/r}$, implying $r/r_g = (c/v_K)^2$.
Further expressing the radial velocity $v_r$ as $v_r = f v_K$, one obtains
\begin{equation}
\frac{\dot{M}}{\dot{M}_{\mathrm{Edd}}} = \frac{\eta}{16}\, f^2 \left(\frac{\tau}{\Theta_p}\right)^2 \left(\frac{m_e}{m_p}\right)^2.
\label{eq:final1}    
\end{equation}
We point out that although no explicit radial dependence exists in Equation \ref{eq:final1}, the result has a radial dependence via the  proton temperature profile,  $\Theta_p = \Theta_p(r)$, and possibly radial variation of $f = v_r/v_K$. Similarly, Equation \ref{eq:final1} should be regarded as an order-of-magnitude estimate, since the critical accretion rate scales as $\dot{M}_{\rm crit} \propto \tau^2$. The characteristic electron--proton coupling parameter ($\tau$) reported in the literature spans $\tau \sim 3$--10, implying an uncertainty of approximately one order of magnitude in the absolute normalization of $\dot{M}_{\rm crit}$. This uncertainty affects the location of the $\dot{M}_{\rm crit}$ but does not alter the underlying scaling with the plasma parameters.

Analyzing the results of S-4 revealed that filamentation, seen in Figure \ref{fig:funnels} becomes noticeable at radii smaller than $\sim 30 r_g$. This is supported by the results presented in Figure \ref{fig:disk-height} which indicate that the disk height of S-4 substantially decreases compared to the numerical results of disks with lower mass accretion rate around that radius. We interpret it as the onset of thinning of filamentation for this cooling rate and chose this radius for our analysis.

\begin{figure}
    \centering
    \includegraphics[width=0.5\linewidth]{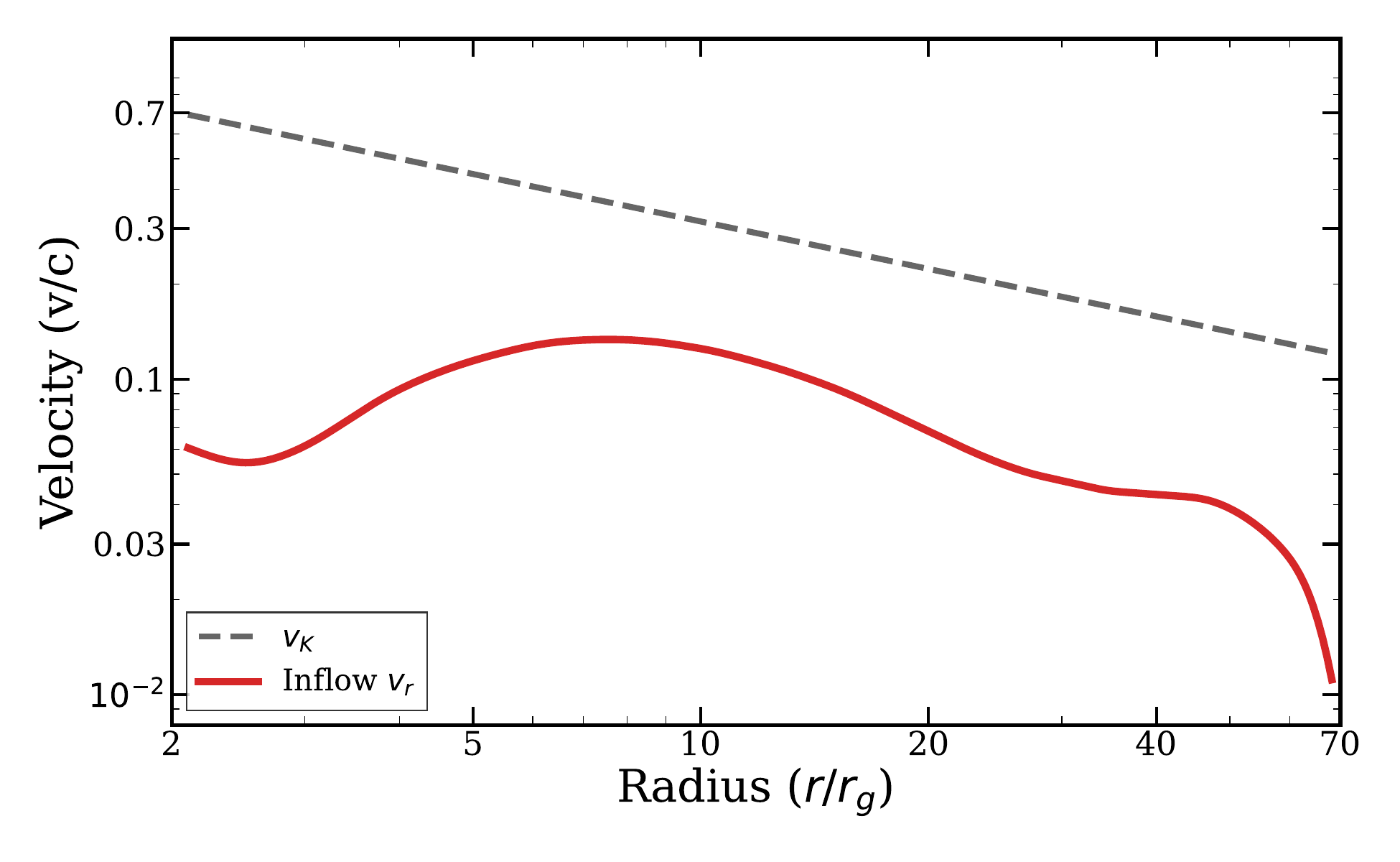}
    \caption{ The radial profile of the inflow velocity ($v_r/c$) shown in the red solid line for an accretion rate of $10^{-4}\,\dot{M}_{\rm Edd}$. The Keplerian velocity ($v_K \propto r^{-1/2}$) is shown in the dashed line for an easy comparison.}
    \label{fig:gas_pressure}
\end{figure}

Using representative values at $r = 30 ~r_g$ for
the radial velocity $v_r = 0.05c$ in Equation \ref{eq:final} (or alternatively, $f \approx 0.3$ in Equation \ref{eq:final1}; see Figure \ref{fig:gas_pressure}) and temperature $\Theta_p = 2.5 \times 10^{-3}$, and assuming $\eta = 0.1$ and $\tau = 3$
for consistency with our underlying assumptions, we find that the cooling rate is comparable
to the heating rate at accretion rate $\dot{M} \sim 2.17 \times 10^{-4}\,\dot{M}_{\mathrm{Edd}}$.
This estimate is consistent with the mass accretion rate at which significant structural
changes are observed in our simulations, supporting the interpretation that radiative
cooling becomes dynamically important within $r \lesssim 30 r_g$ in the S-4 model.

At lower accretion rates, cooling is not efficient enough to significantly reduce the gas temperature. As a result, the gas remains hot and approximately in equipartition with the magnetic energy density. This balance allows the filaments to sustain their vertical pressure support, thereby maintaining a relatively constant and puffed up disk height across these lower accretion rates, as is demonstrated in Figure \ref{fig:funnels}.
Consistent with this expectation, we observe that the disk height at $10^{-4}\,\dot{M}_{\mathrm{Edd}}$ shows a decline compared to the lower accretion rate cases where it remains nearly unaffected by the cooling. 

\section{Discussion and Conclusions}
\label{sec:discussion}

In this work, we explored the effects of radiative cooling on the structure and dynamics of accretion flows around non-spinning BH. Recent two-temperature GRMHD simulations of magnetically arrested disks have demonstrated that cooling processes can noticeably modify the electron thermodynamics. In particular, synchrotron emission reduces the electron temperature in the inner disk and leads to a measurable ($\sim 10\%$) suppression of the (sub)millimetre synchrotron flux and its temporal
fluctuations, even for $\dot{M} \sim 10^{-7}\dot{M}_{\mathrm{Edd}}$ \citep{Salas+24}. This effect arises because the synchrotron cooling timescale is only a few times longer than the accretion timescale, with $\tau_{\mathrm{sync}} / \tau_{\mathrm{accr}} \sim 3$, indicating that cooling cannot be neglected in the regime of low mass accretion rate \citep{Salas+24}. 

To study the effects of radiative cooling, we
performed a series of GRMHD simulations of accretion disks in the MAD regime around a non rotating BH. We included the radiative cooling via synchrotron and bremsstrahlung processes inside the simulation. Compton cooling was exempt from our calculation, as it requires tracing the local radiation field. But this process was shown to be negligible at the mass accretion rates considered here. 
Similar to several previous works, in our calculations we used the single temperature prescription, namely a fixed ions to electrons temperature ratio, $T_i/T_e$. We selected here a value of $T_i/T_e = 3$, which is within the range discussed in the literature \citep{Dibi+12,Drappaeu+13, Yoon+20}.
A set of simulations were performed, probing accretion rates in the range $10^{-7} $-- $10^{-4} \dot{M}_{\rm Edd}$. We had previously ran a complementary non-cooling simulation to normalize the measured accretion rates and to provide a benchmark for evaluating the effects of cooling.

The key results of our work are as follows:

\begin{enumerate}
\item We find that the traditional definition of disk height (Equation \ref{eq:old-disk-height}) does not accurately represent the disk thickness, $h/r$, because it inherently assumes that the density peaks at the disk midplane. However, the simulations we present here, as well as many previous simulations of disks in the MAD state, clearly show that this assumption does not hold. In the innermost regions of the disk, the gas flow becomes strongly filamented due to interchange instability, and these filaments  penetrate the magnetosphere and pave their way towards the BH in a non-symmetric path (see Figure  \ref{fig:funnels}).
Under these highly dynamical and non-axisymmetric conditions, the midplane density often becomes a poor proxy for the bulk of the matter distribution. Relying on a midplane-centered integration leads to an artificial increase of the physical scale height, as it fails to account for the vertical displacement of these high-density filamentary structures. 

To address this, we propose a new definition of the disk height, based on the location of the density maxima (Equation \ref{eq:new-height}). By tracking the actual mass concentration rather than a fixed geometric plane, the vertical extent of the accreting gas is captured more faithfully. Specifically, our method identifies filaments (density peaks) on an azimuthal basis, allowing the diagnostic to follow the local path of the filaments rather than averaging over the non-symmetric voids inherent in the MAD state. Using this new definition, we find that the disk height approaches zero on the BH horizon (see Figure \ref{fig:disk-height}).

\item
Our results show that the disk height decreases as the rate of radiative cooling increases. We identified, both numerically
and analytically, a critical mass accretion rate above which radiative cooling becomes substantial enough to have a notable
effect on the structure and scale height of the accretion disk beyond some prescribed radius. Once this critical
accretion rate is reached, the disk height exhibits a sharp decline due to the significant loss of thermal pressure caused
by enhanced radiative cooling.

%\item 
In Section \ref{sec:cooling-derivation} we derived an analytical expression for this critical mass accretion rate. This
derivation is based on the assumption that 
in the inner region of the disk
there is a strong coupling between the magnetic
field and the gas thermal pressure. 
We showed that when the mass accretion rate is $\sim 10^{-4}\dot{M}_{\text{Edd}}$, the
radiative cooling becomes efficient in radiating away the thermal energy from the inner disk region ($\lesssim 40 r_g$ ) faster than it can be replenished.
This shift in the energy balance modifies the internal disk structure and leads to a significant decline in the disk thickness as the gas loses the pressure support needed to remain puffed up. 
\end{enumerate} 

We further observe that the local cooling rate $q^{-}$ decreases monotonically with increasing radius, reflecting the fact that most radiative losses occur in the innermost disk regions (see Figure \ref{Cooling components}). In these regions, the magnetic field, as well as gravitational energy release, gas temperature, and particle density are highest, enhancing synchrotron cooling. 
We also find that synchrotron cooling is the dominant radiative mechanism over Bremsstrahlung at all mass accretion rates ($10^{-7} $-- $10^{-4} \dot{M}_{\rm Edd}$) considered in this work. Bremsstrahlung cooling becomes more noticeable at large radii but remains subdominant overall by several orders of magnitude.

The agreement between our analytical estimate and the numerical results indicates that our
model successfully captures the key physical ingredients of radiatively cooled MADs. 
Based upon the assumption of strong coupling between the gas and the magnetic field, this
derivation complements and adds to our previous work \citep{Singh25}. In that work, we
considered the gas energy balance with gain from accretion and radiative losses, along with
the saturation of the MAD parameter as a key factor, and found a critical mass accretion rate
(normalized to Eddington accretion rate) which is independent on the BH mass. Here, we
replaced the assumption of saturated MAD parameter by the assumption of a strong coupling
between the gas and the magnetic field in the inner disk regions. This enabled us to
assert the effects of gas cooling at large distance from the horizon. The newly calculated
critical mass accretion rate depends on the radius up to which cooling is efficient, see
Equation \eqref{eq:final}.  

The present study is restricted to a non-spinning black hole in order to isolate the effects of radiative cooling on the MAD structure. Black hole spin is expected to modify the saturation of magnetic flux, the inner disk geometry, and the efficiency of jet launching, which may shift the normalization of the critical accretion rate. However, the cooling-driven transition identified here is expected to remain qualitatively unchanged. A systematic exploration of spin dependence will be presented in future work.

A key assumption in our analysis is a fixed proton-to-electron temperature ratio, $\tau \equiv T_p / T_e = 3$.
This choice is well motivated in the regime of low accretion rates, $\dot{M} \lesssim 10^{-6}\dot{M}_{\mathrm{Edd}}$,
where previous studies have shown that $\tau$ remains relatively stable throughout the flow
\citep[e.g.][]{sadowski+17, Yoon+20}. Even as the accretion rate increases to
$\dot{M} \sim 10^{-4}\dot{M}_{\mathrm{Edd}}$, deviations from this value are modest \citep{sadowski+17}.
At higher accretion rates, however, the assumption of a constant temperature ratio becomes less secure.
In this regime, radiative cooling can dominate over Coulomb coupling, leading to a more complex and
model-dependent electron-proton energy partition. Different prescriptions for the electron to proton
temperature ratio capture this behavior in distinct ways, see e.g. \citet{Rowan+19,Kawazura+19}. A systematic exploration
of how different electron--proton coupling models impact our conclusions at higher accretion rates is deferred to future work.

Our simulations demonstrate the effect of radiative cooling on the structure of MADs. These disks are
not made of a continuous fluid, but rather a multi-phase environment where high-density filaments move
through a low-density, magnetically dominated background. As asserted by our analysis, the thickness
of these filaments strongly depends on the radiative cooling, and therefore on the mass accretion rate.
We further speculate that their dynamics may also be modified by radiative cooling. Future models of
spectral energy distributions (SEDs) and fluctuations in the light-curves must account for these
non-symmetric structures, as the localized synchrotron hotspots within these filaments may produce
different variability patterns depending on the mass accretion rate. This could be probed by
ray-tracing our simulations.

We expect that the dominance of synchrotron cooling over bremsstrahlung, see Figure
\ref{Cooling components}, could have major implications for the appearance of cooled
MADs. In the inner region, the disk structure is uniquely sensitive to the magnetic
field strength, which is not just a driver of accretion, but the primary regulator
of how fast the disk cools.
Consequently, provided that changing-look AGNs or X-ray binaries are powered by
high-accretion MADs \citep[see e.g.][]{You+23}, radiative cooling should imprint itself on the inner disk dynamics,
and as such on the observational signatures of such a system. This might provide an
independent way to measure mass accretion rate. 

\section{acknowledgement}
\begin{acknowledgments}
We thank the referee for the constructive comments and valuable suggestions, which helped improve this paper. The Research required numerical calculations that were carried out on the NVIDIA Israel-1 supercomputer, using time generously allocated to us by NVIDIA. The NVIDIA Israel-1 system which includes NVIDIA H100 GPUs and NVIDIA Spectrum-X infrastructure (NVIDIA Spectrum-4 Switches and NVIDIA BlueField-3 SuperNICs), allowed the model to gain a performance improvement of 4, vs BIU node of 4 A100 GPUs. 
\end{acknowledgments}

\bibliography{ms}{}
\bibliographystyle{aasjournal}
\end{document}